                               \documentstyle[12pt,aasms4]{article}
\def\gtrsim{\mathrel{\hbox{\rlap{\hbox{\lower4pt\hbox{$\sim$}}}\hbox{$>$}}}}
\let\ga=\gtrsim
\def\lesssim{\mathrel{\hbox{\rlap{\hbox{\lower4pt\hbox{$\sim$}}}\hbox{$<$}}}}

\begin{document}

\title{Chandra Analysis of Abell 496 - No Chemical Gradients Across Cold Fronts}
\author{Renato Dupke}
\affil{University of Michigan, Ann Arbor}

\author{Raymond E. White III}
\affil{University of Alabama, Tuscaloosa}

\begin{abstract}

We present the results of a spatially-resolved spectroscopic analysis of the 
galaxy cluster Abell 496 with the S3 chip on-board the {\sl Chandra} satellite. 
We confirm the presence of a central positive temperature gradient consistent with a
cooling flow, but with a minimum gas temperature of $\sim$0.5--0.9 keV.
The cluster also exhibits sharp edges in gas density and temperature which are
consistent with ``cold front'' substructures. 
The iron abundance profile is not radially symmetric relative to the cluster center.
Towards the direction of the most prominent (northerly) cold front, the 
iron abundance is roughly flat, with nearly solar values. 
In the opposite (southerly) direction from the center, the iron abundance distribution
shows an ``off-center'' peak. 
Various abundance ratios suggest that the heavy elements in the central regions
of the cluster are dominated by SN Ia ejecta. 
However, for radii greater than 100 $h_{50}^{-1}$ kpc, the abundance ratios 
vary in such a way that different abundance ratios provide very different
estimates of the proportion of SN Ia/II ejecta.
Nonetheless, observed abundances and abundance ratios are continuous across the cold fronts,
which suggests that the cold fronts are not likely to be the result of a subcluster merger.
We suggest instead that the cold fronts in A496 are caused by ``sloshing" of the central
cooling flow gas, induced by the motion of the cD about the cluster center.

\end{abstract}


\section{Introduction}

{\sl Chandra} satellite observations have revealed a wealth of X-ray structures 
in the cores of galaxy clusters,
including X-ray plumes (e.g. Sanders \& Fabian 2002), cavities (e.g. McNamara et al.\ 2000),
and ``cold fronts''
(e.g.\  Markevitch et al.\ 2000; Vikhlinin et al.\ 2001; Mazzotta et al.\ 2001). 
Cold fronts (hereafter CF) are sharp surface 
brightness discontinuities (widths are typically smaller 
than the electron mean free path) characterized by a jump in gas temperature, accompanied by
a fall in X-ray surface 
brightness such that the gas pressure remains continuous across the front. 
Thus, these structures differ from
bow shocks (e.g. Markevitch et al.\ 2002) and are often attributed to 
subsonic (transonic) motions of accreted substructures 
(Markevitch et al.\ 2000, 2001; Vikhlinin et al.\ 2001)

The incidence of gaseous substructures at the centers of clusters suggests that cluster
cores are very dynamic, even when there are no obvious 
signs of merging (such as in cooling flow clusters with 
regular isophotes). In this {\it Letter} we show that this is the case for A496. 
This cluster is a typical, bright, nearby (z$\approx$0.032), 
apparently well-behaved cooling 
flow cluster. In a joint {\sl Ginga} and  {\sl Einstein} analysis of A496, White et al.\ (1994) found evidence
for a central metal abundance enhancement. This was corroborated 
by {\sl ASCA} (e.g. Dupke \& White 2000a), {\sl SAX} 
(Irwin \& Bregman 2001) and {\sl XMM} (Tamura et al.\ 2001) with greater statistical significance. 
Furthermore, Dupke \& White (2000a) also discovered radial gradients in various 
 abundance $ratios$, indicating that the gas in the central 2-3$^\prime$ has 
a higher proportion of SN Ia ejecta ($\sim$70\%) than the outer parts of the cluster. 

Spectral analysis of 
radially averaged 
annuli with {\sl XMM} 
is roughly consistent with 
{\sl ASCA} results (Tamura et al.\ 2001).
The spatial resolution achieved by
the instruments on-board {\sl XMM} are very limited in establishing fine correlations 
between spatial substructures,
such as CFs, and spectral parameters such as metal abundances and temperatures.
However, this analysis  is 
important since it provides 
clues about the nature of such
structures. For example, if CFs are caused by the passage of a high velocity gas clump 
through the 
intracluster gas, the sharp surface discontinuity should be accompanied by 
chemical discontinuities as well, representing the enrichment history of the two
different systems.
This kind of analysis can be 
performed best with {\sl Chandra}, given its high angular resolution. 
In this {\it Letter} we describe the discovery of CFs in A496 and search
for chemical discontinuities across them. 
A more detailed analysis of the gas mass distribution
will be published elsewhere. All distances shown 
are calculated assuming a H$_0=50$ km~s$^{-1}$Mpc$^{-1}$ and $\Omega_0=1$. At the distance of this 
cluster $1^{\prime\prime}\approx$ 0.95 kpc.

\section{Data Reduction}

A496 was observed by {\sl Chandra} ACIS-S3 in July 2000 and October 2001, for 20 
and 10 ksec, respectively. In both cases the cluster was centered on the S3 chip. 
The first observation 
was largely contaminated by flares.
and will not be analyzed here. 
We used Ciao 2.2.1 with CALDB 2.9 to screen the data.
Two short flare-like periods were extracted and the resulting exposure time 
is 8.7 ksec. A gain map correction was applied together with PHA and pixel randomization. 
The ACIS particle background was cleaned as prescribed for VFAINT  
mode. Point sources were extracted and the background used
in spectral fits was generated from blank-sky observations using
the {\tt acis\_bkgrnd\_lookup} script. 
Here we show the results of spectral fits with XSPEC V11.2 (Arnaud 1996) using the 
{\tt mekal} and {\tt vmekal} thermal emission models. 
An isobaric cooling flow model {\tt cflow}, was used for the spectral fittings of 
the inner regions, where the minimum cooling flow temperature ($kT^{\rm cf}_{\rm min}$) was  
allowed to vary.
Metal abundances are measured relative to the solar photospheric values of 
Anders \& Grevesse (1989).
Galactic photoelectric absorption was incorporated using the {\tt wabs} 
model (Morrison \& McCammon  1983).
Spectral channels were grouped ($>$25 cnt/chan). Energy 
ranges were restricted to 0.5--8.5 keV. 
Errors are 1-$\sigma$ unless stated otherwise.

In order to compensate for the recently detected degradation of ACIS low energy quantum efficiency (QE), 
we used the most recent {\tt corrarf} routine, which applies the absorption model {\tt acisabs} 
(Chartas \& Getman (2002))
to the effective area file generated by the CIAO tool {\tt mkarf}. The resulting spectral fits have 
significantly reduced column densities (compared to the prior, artificially inflated values), 
but they have a mild dependence on the low energy cut-off used in the data. 
Low-energy metal abundances 
are also affected by the low energy cut-off value. Since we are looking for {\sl relative} changes
of spectral parameters through different directions inside the cluster, this problem does
not affect our conclusions. However, the absolute values of some  abundances and
of the hydrogen column density should not be considered definitive since the low energy 
calibration is still evolving.

\section{Results}
\subsection{Cold Front}

Figs.~1a and b show the X-ray image of A496. One can clearly see the sharp edge in
surface brightness (SB) towards the north.
Fig.~1a shows the directions towards {\it sb\_sharp} and {\it sb\_smooth} used to 
extract SB profiles. We also extracted SB profiles for the 
perpendicular directions (labeled {\it sb\_east} and {\it sb\_west}) for comparison.
Fig.~1b shows the extraction 
regions used for spectral analysis, which are separated in semi-annuli with opening angles close
to 180$^\circ$ towards the {\it sharp} direction  and towards the opposite 
direction (hereafter called {\it smooth}). The semi-annuli were chosen to maximize photon statistics at
several radii along the direction of interest (perpendicular to the SB edge). 

Fig.~2 shows a comparison between the SB and temperature profiles. 
The temperatures were typically obtained through spectral fits using 
an absorbed {\tt mekal} model, where the hydrogen column density was a free parameter. 
Despite the apparent smoothness of the average SB profile (in black) it is clear 
that the density profiles towards different directions
are irregular, especially along the {\it sb\_sharp}--{\it sb\_smooth} axis. 
The position of the sharp edge is shown by a red vertical red line (at $r\approx$ 82 kpc), where 
there is a 
strong decline in SB: a factor of $\sim$2 within $\sim$10 kpc, accompanied by a temperature 
rise from $4.06_{-0.28}^{+0.26}$ keV to 5.20$_{-0.42}^{+0.46}$ keV. 
The gas pressure is 
roughly constant (within the errors) throughout 
the discontinuity,  which is consistent with the CF
phenomenon (Markevitch et al.\ 2000). The temperature profile within 
82 kpc also exhibit strong anisotropies. Towards the {\it sharp} direction 
the temperature rises more or less 
steadily from 1.66$_{-0.11}^{+0.13}$ to 4.06$_{-0.28}^{+0.26}$ keV. 
The addition of a cooling flow model in the 
inner 10$^{\prime\prime}$  does not change significantly the best-fit parameters and does not improve
the $\chi_{\nu}^2$. 
The minimum cooling flow temperature found is $kT^{\rm cf}_{\rm min}~\sim$0.96 keV,
with an associated mass deposition rate of $\sim$5 $M_\odot$ yr$^{-1}$.
Towards the {\it smooth} side the temperature rises steadily outward to a radial distance 
of 25 kpc and then rises steeply from
2.73 $_{-0.21}^{+0.25}$ keV to 4.30 $_{-0.42}^{+0.48}$ keV within a span of 15 kpc. 
The temperature profile
suggests the presence of a secondary CF at $\sim$ 25 kpc, although there is no clear association
with a SB drop within the same region.
This may be due to the azimuthally symmetric binning geometry being a 
poor match to the oblique front structure to the south.
At r $\ga$ 25  kpc the gas temperature has a flat profile throughout. 
Spectral fittings of the central region on the {\it smooth} side are significantly improved
with the addition of a cooling flow component. 
The resulting best fit temperature is 3.2$\pm$ 0.7 keV with a mass deposition 
rate of 4.6$\pm$1.3 $M_\odot$ yr$^{-1}$ and
 $kT^{\rm cf}_{\rm min}$ = 0.57$_{-0.49}^{+0.25}$ keV.

\subsection{Metal Abundance and Abundance Ratios Distribution}

The spectral fittings when individual  abundances were allowed to vary independently 
({\tt vmekal}) were 
slightly better than those where the abundances were tied to their solar ratios, as previously found 
in {\sl ASCA} data by Dupke \& White (2000a).
Fig.~3 shows temperature and abundance profiles of the best constrained elements (Fe, Si, 
O and S) in radial slices through the whole region analyzed in this work.
Temperature and
iron abundance ($A_{\rm Fe}$) anisotropies are clearly seen. The radial binning in Fig.~3 is larger than that 
in Fig.~2. This choice of
binning increased the significance and magnitude of the temperature jump across the CF. 
The temperature jumps from 3.57$_{-0.20}^{+0.23}$ keV to 
5.60$_{-0.32}^{+0.51}$ keV across the CF. Note that $A_{\rm Fe}$ does not change 
across the CF.
The $A_{\rm Fe}$ profile is remarkably flat towards the {\it sharp} direction, with an 
average value of 
$\sim$0.75 solar. Towards the 
{\it smooth} direction $A_{\rm Fe}$ has a marginal rise from 0.59$_{-0.10}^{+0.17}$ solar at the center to 
0.87$_{-0.15}^{+0.17}$ solar at 
$\sim$ 70 kpc. It then declines again to a minimum value of 0.49$_{-0.08}^{+0.10}$ solar towards 
the outer regions.
We used the F-test to estimate the significance of the Fe abundance gradient towards the {\it smooth}
side. 
We verified that the abundance gradient in the fourth and fifth radial bins is significant at $\ga$ 98\% confidence.
This off-center abundance peak is similar to that found by Sanders \& Fabian (2002) in A3526 and by 
Johnstone et al.\ (2002) in A2199. 
However, 
the addition of a cooling flow component to the innermost spatial bin ($<$22$^{\prime\prime}$)
improves the spectral fits
significantly, increasing the best fit $A_{\rm Fe}$ 
to 1.52$\pm0.45$ solar, eliminating the central depression.
Adding the same cooling flow component
to the {\it sharp} central region does not change significantly $A_{\rm Fe}$. 

The distribution of the silicon abundance ($A_{\rm Si}$)
shows a negative gradient towards both {\it sharp}
and {\it smooth} sides. The gradient is steeper but consistent with that measured in full (0-2$\pi$) 
annular regions 
with {\sl XMM} data (Tamura et al.\ 2001). The oxygen abundance ($A_{O}$) tends to rise radially towards both 
{\it sharp} and {\it smooth} directions. By simultaneously fitting  the spectra of 
the inner bins from the {\it sharp}
and {\it smooth} sides, we measured a best fit central $A_{O}$ 
of 0.60$_{-0.36}^{+0.24}$ solar, which is significantly lower than in the 
outermost bin of the {\it smooth} side ($A_O=1.61_{-0.45}^{+0.6}$ solar). 
This trend is consistent with {\sl ASCA} SIS 
measurements (Dupke \& White 2000a) but was not observed by Tamura et al.\ (2001). However,
the sizes of the extraction regions used in their work are not directly comparable 
to ours. Furthermore, as it can be seen from Fig.~3, their choice of region configuration can smooth out the
$A_{O}$ gradient 
due to the asymmetries in the abundance profile.
The central absolute $A_{O}$
values are higher than those found in a preliminary {\sl Chandra} 
analysis of Dupke \& White (2001), who used the least flare-contaminated period of a
previous observation of A496 without any correction for QE degradation. 
The sulfur abundance
is not well constrained and is consistent with being flat. 
Overall, $A_{\rm Si}$ and $A_{O}$ distributions are consistent with radial
gradients, but there are no discontinuities across the CFs.

In Fig.~4 we plot the abundance ratio profiles. Since different types of SNe
have characteristic
mass yields for different elements, their relative pollution rate can be measured, in principle,
via abundance ratios (cf.\ Loewenstein \& Mushotzky 1996), which in turn 
help us to determine the enrichment history of the gas. It can be seen that both O/Fe and Si/Fe are 
consistent with a central ($r\lesssim$ 80 kpc) dominance of SN Ia ejecta. 
However, both ratios show radial gradients and their outer values do not agree on
a single proportion of SN Ia/II ejecta at a given exterior radius.
The low Si/Fe ratios between 100--200  kpc are driven mainly by the low values of $A_{\rm Si}$ 
in that region. 
A positive gradient in the O/Fe was also 
detected in Dupke \& White (2001) and it is unlikely that 
further calibration improvements will affect significantly the
O/Fe profile, although the 
absolute values of the individual abundances are more uncertain. 
The poor photon statistics of our observation do not allow us
to constrain other ratios well enough to check for the self-consistency of abundance ratios using
different theoretical SN models, as in  Dupke \& White (2000b). 
However, independently of absolute normalizations, 
it is clear that there is
no significant change in abundance ratio distribution across the sharp boundaries of the CFs.

\section{Discussion: The Nature of Cold Fronts}

Two recently proposed hypotheses for generating CFs in clusters 
involve the infall of galaxy groups into the larger clusters. 
In some cases the substructures are sub(/trans)sonic remnants of mergers 
(e.g. Markevitch et al.\ 2000; Vikhlinin et al.\ 2001). 
In other cases, such as A1795 (Markevitch et al.\ 2001), it is proposed that
the infall of a small gravitational substructure disturbs the 
central gravitational potential 
of the main cluster, causing low-entropy gas in the center to oscillate around
some equilibrium position (the ``sloshing'' hypothesis).
In both cases 
one expects the CF to be a surface discontinuity between a cold core moving
through hotter, diffuse intracluster gas, with inefficient
mixing between the different gas phases across the surface discontinuity. 
If this were true for A496 one should expect to see different metal enrichment 
histories across the CF.  
We find instead that abundances and abundance ratios are continuous
across CFs, a result that is independent of calibration
uncertainties in determining the absolute abundances of individual elements.

The data is roughly consistent with a model where the cD is oscillating 
around the clusters' potential well. This model is similar to that 
suggested for A1795 (Fabian et al.\ 2001). In both clusters the CFs
are aligned with the most likely projected axis of motion of the cD. This axis is 
inferred by analysis of 
the direction of the main extension of optical filaments (see Fabian et al.\ 1981),
X-ray and optical isophotal elongation. The cDs in both systems have similar 
peculiar velocities after correcting for substructure (Bird 1994).  In this model 
the cD would be dragging/smearing
SN Ia Fe enriched cold gas within the spatial oscillation length ($\sim$ 80 kpc), which 
defines the distance to farthest CF observed. The most likely evolutionary time frame
for this model would place the cD in A496 moving south beginning to generate a 
second CF. 
The motion of the cD may add to other competing core heating mechanisms, which may keep 
$kT^{\rm cf}_{\rm min}$ $\sim1$ keV.
A detailed analysis of this model and other alternates is beyond 
the scope of this {\it Letter} and will be published elsewhere. 
The investigation of the behavior of abundances and their 
ratios across SB discontinuities in the cores of other clusters
will allow us to determine whether such structures are typically induced externally, via
subcluster mergers, or internally, perhaps through sloshing associated with cD motion.

\acknowledgments 
We are very grateful to P. Plucinsky, A. Prestwitch and 
H. Tananbaum for their help in obtaining a reobservation of A496. 
We acknowledge support from NASA through {\sl Chandra} award number 
GO 0-1085X, issued by the {\sl Chandra} X-ray Observatory Center, which is
operated by the Smithsonian Astrophysical Observatory for and on behalf of
NASA under contract NAS8-39073.  RAD was also partially supported by 
NASA grant NAG 5-3247.  RAD also thanks J. Irwin,
J. Bregman and E. Lloyd-Davies for helpful discussions.

\clearpage
                                \title{
Figure Captions
                                }
				\figcaption{
A496 Extraction Regions. X-ray exposure corrected image (in the range 0.3--10 keV).
 One pixel corresponds to 4$^{\prime\prime}$ and North is to the top. (a) The four different
 directions selected for image analysis. (b) Semi-annuli extraction regions
  used in the spectral analysis.
                                }

                                \figcaption{
Temperature and Density Profiles. {\it Top}: Best-fit gas temperature distributions towards the 
{\it sharp} (red) and {\it smooth} 
(blue) directions obtained using a {\tt wabs mekal} spectral model and also with the addition
of a {\tt cflow} component for the central bin (points  with vertical error bars only). 
The extraction regions correspond to those shown in Fig.~1b. Errors
are 1$-\sigma$ confidence. Vertical red line indicates the position of the sharp edge 
towards the {\it sharp} direction. {\it Middle}: Surface brightness profiles towards different 
directions indicated in Fig.~1a. Similar color notation is applied for the 
{\it sb\_sharp} and {\it sb\_smooth} directions. {\it East} and {\it west} directions 
are also shown in green and dark yellow colors, respectively. Splines connecting surface 
brightness values for all directions are also shown (with normalizations reduced by 
a factor of 2 for illustration purposes). We also show the overall (full annuli) 
surface brightness profile (black line with normalization reduced 
by a factor of 4). {\it Bottom}: Reduced chi-squared for the spectral fittings shown on 
{\it Top} plot, using the same color notation. Spectral fittings typically have 50--100 degrees 
of freedom. Horizontal lines show the values for the cases where a {\tt cflow} 
spectral component is added in the spectral fits for the central bin. 
}

				\figcaption{
Individual Metal Abundance \& Temperature Distributions. Results from spectral fittings of 
regions along the 
line of symmetry of the sharp edge. The zero point represent the X-ray center. 
The vertical
line indicates the position of the CF. Metal abundances are denoted by the element notation. 
The bin size used for Si, O and S is larger than that 
of T and Fe to improve
statistics. Fittings with an additional {\tt cflow} component to the {\tt wabs Vmekal} are also 
shown for the central 
bin in all plots by points with vertical error bars only. The reduced chi-squared for the spectral 
fittings are shown 
on the bottom plot for all bin sizes (black for T, Fe and gray for Si, O, S).  
Spectral fittings typically have 100--200 and
150-- 250 degrees of freedom for thin and thick bins, respectively. Horizontal 
lines show the values for the cases where a {\tt cflow} component is added to the spectral
models used to fit data from the central 
bin (black solid lines for T, Fe and gray dashed lines for Si, O, S).
                               }

				\figcaption{
Individual Metal Abundance Ratio Distributions. Notation is analogous to Fig.~3. 
Dotted and dashed horizontal 
lines show the predicted values for 100\% SN Ia contamination and 100\% SN II contamination, 
respectively, as described by Nomoto et al.\ 1997a,b. Central bin size is 37.5$^{\prime\prime}$.
				}

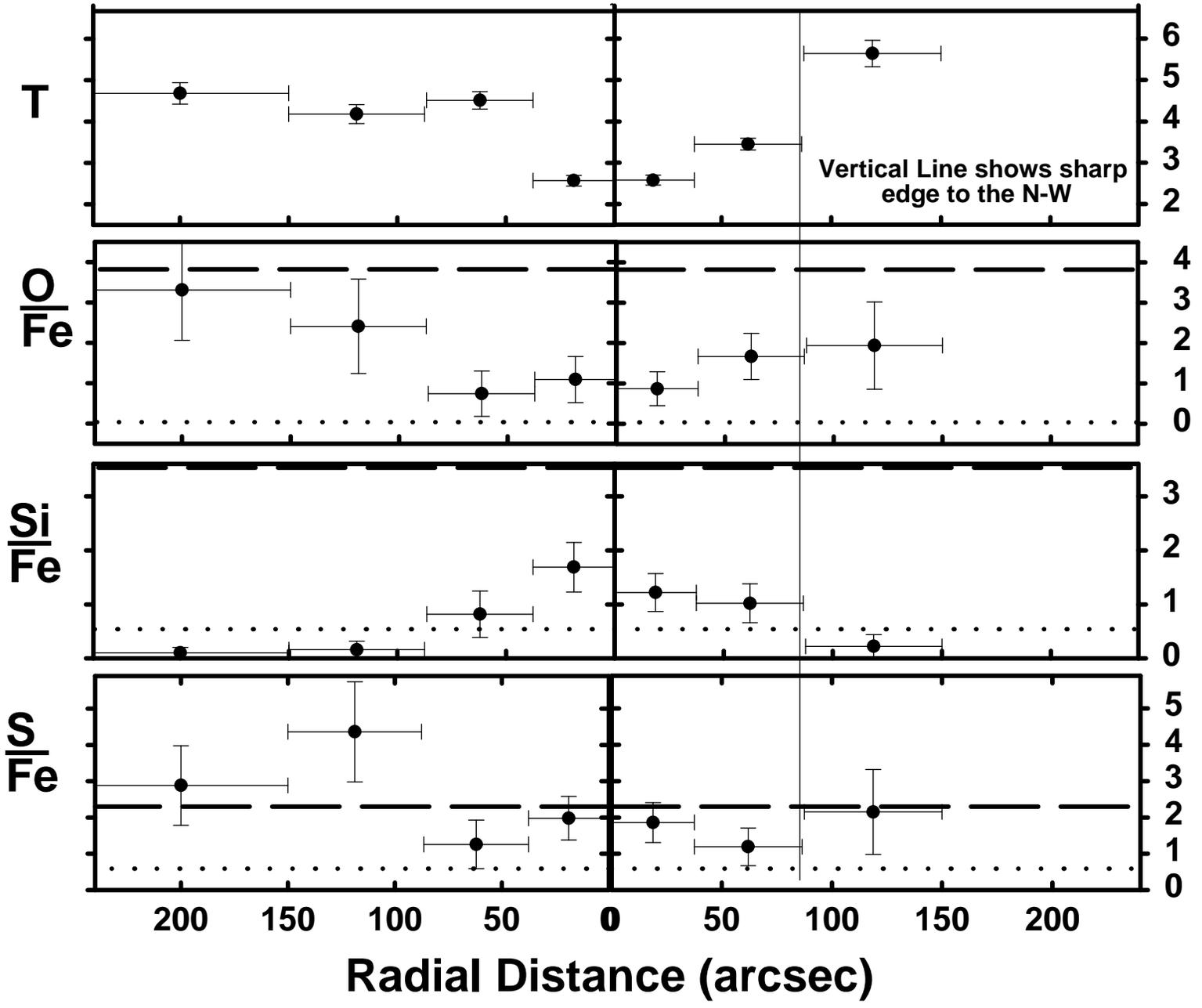
\end{document}